\documentstyle[prl,aps]{revtex}

\begin{document}

\title{ Effective Potential, Mori Product and Quantum Dynamics}
\author{A.Cuccoli$^{1,2}$, R.Giachetti$^{1,3}$, V.Tognetti$^{1,2}$ and
  R.Vaia$^{2,4}$} \address{$^1$Dipartimento di Fisica
  dell'Universit\`a di Firenze.  \\$^2$Istituto Nazionale di Fisica
  della Materia (INFM), Unit\'a di Firenze.  \\$^3$Istituto Nazionale
  di Fisica Nucleare (INFN), Sezione di Firenze.  \\ $^4$ Istituto di
  Elettronica Quantistica del C.N.R. Firenze}

\date{\today}

\maketitle

\begin{abstract}
  We present a method that permits the calculation of the dynamical
  correlation functions for quantum systems. These are obtained by
  evaluating the generating functionals of the static moments of the
  relaxation functions in a self-consistent approximation that can be
  obtained by an appropriate extension of the effective potential
  theory.
\end{abstract}


\vskip 10 truemm
\widetext

The thermodynamics of quantum systems has been widely studied by the
effective potential theory \cite{GTall,FeynmanK86}. This treatment has
proved to be very accurate to determine equilibrium properties
\cite{CTVVmagall,CowleyH95,CGTVV95}, but encounters quite a number of
theoretical and practical difficulties in being applied to dynamical
phenomena. An extension of the formalism uniquely based on ingenuity
does not allow for a serious control of the results and slips easily
in {\it ad hoc} or even erroneous assumptions. On the other hand it is
self-evident that an adequate way for calculating dynamical
correlations would be highly welcome.  It is our purpose to try to
fill this gap and to provide an effective method to approach dynamical
calculations in terms of path integrals.  We shall mainly be concerned
with the Kubo relaxation functions, that naturally appear in the
framework of Mori theory \cite{Moriboth}. These functions are obtained
by suitably defined scalar products, the Mori products
\begin{equation}
 R_{A,B}(t)=(\hat A|\hat B(t)) =
 \int_0^{\beta\hbar}\,du\,\langle\,\hat A(0)\,\hat B(t+iu)\,\rangle\,,\
 \label{1}
\end{equation}
where the observables $\hat{A}$ and $\hat{B}$ are taken such that
$\langle\hat{A}\rangle=\langle\hat{B}\rangle=0$, and
braces denote the thermodynamic average. The Laplace
transform of Eq.(\ref{1}), in turn, has a continued fraction
representation whose coefficients are static normalized relaxation
functions, {\it i.e.}, time-independent normalized Mori products, of
derivatives of the dynamical variables $\hat A(t)$ and $\hat B(t)$. In
particular, the self-relaxation function $\Xi_0(t)=(\hat F_0|\hat
F_0)^{-1}\,(\hat F_0(t)|\hat F_0)$ of an hermitian operator $\hat
F_0(t)$ can be Laplace transformed and expanded in a continued
fraction, namely $\Xi_j(z)=(z+\delta_{j+1}\Xi_{j+1}(z))^{-1}$, where
$\delta_{j+1}=(\hat F_j|\hat F_j)^{-1}\,(\hat F_{j+1}|\hat F_{j+1})$
and $\hat F_j$ denotes the so-called $j$-th fluctuating force. The
quantities $(\hat F_j|\hat F_j)$, can be related to a combination of
the first $2(j+1)$ moments of the time series-expansion of $\Xi_0(t)$.
While $(\hat F_j|\hat F_j)$ with $j\not=0$ can be expressed in terms
of static correlations, the quantity $(\hat F_0|\hat F_0)$ requires
the direct evaluation of the Mori product.

Experiments measure the spectral shape, given by the Fourier transform
of the dynamic correlation function, related to $\Xi_0(z)$ by the
``detailed balance'' principle:
\begin{equation}
 {\cal S}(\omega)={(\hat F_0|\hat F_0)}\,\frac\omega{1-e^{-\beta
 \omega}}\,\frac 1\pi \,\Re\, (\Xi_0(z=i\omega)).
\end{equation}
Therefore, the calculation of ${\cal S}(\omega)$ can be approached from
the knowledge of the static quantities $\delta_j$ up to a sufficiently
large number $j=J$ \cite{CMGTVall}, supported by some insight into the
long-time behaviour of the continued fraction termination $\Xi_J(t)$
\cite{CGTVV95,LoveseyM72,CMGTVall}.

We shall just provide an explicit formula for the {\it direct}
evaluation of an arbitrary $\delta_j$, $(\hat F_0|\hat F_0)$ included,
by an effective potential approximation. We start from the generating
functional in the hamiltonian path-integral form:
\begin{eqnarray}
Z[L,J] &=& \oint{\cal D}[x(u)]\,\int{\cal
D}[p(u)]\,\cr
\exp\Bigl[&-&\frac 1\hbar\,\int_0^{\beta\hbar}\,du\Bigl
(-ip(u)\dot x(u)+{\cal H}\big(p(u),x(u)\big)-\hbar L(u)p(u) - \hbar
J(u)x(u)\,\Bigl)\Bigr]\,.
\end{eqnarray}

According to the effective potential method
\cite{GTall,FeynmanK86,CGTVV95} we consider a quadratic trial action
obtained by substituting ${\cal H}(p(u),x(u))$ with
$(1/2m)\,(p(u)-\eta)^2+(m\omega^2/2)\,(x(u)-\xi)^2+w$, whose
parameters $w$, $m$ and $\omega$ depend on the average point
$(\eta,\xi)= (\beta\hbar)^{-1}\big(\int du\,p(u),\int du\,x(u)\big)$
of each path. The effective Hamiltonian reads:
\begin{equation}
  {\cal H}_{{}_{\rm
      eff}}(\eta,\xi)=w(\eta,\xi)+\beta^{-1}\,\ln(f^{-1}\,\sinh f)~,
\end{equation}
where $f=\beta\hbar\omega/2$ rules the quantum character of the
system. By defining the two-component vectors $\rho={}^t(\eta,\xi)$
and $K(u)={}^t(L(u),J(u))$, the approximated generating functional can
be written as:
\begin{equation}
  Z_0[K] =\int
  \frac{d\eta\,d\xi}{2\pi\hbar}\,e^{\displaystyle{-\beta{\cal
        H}_{{}_{\rm eff}}(\eta,\xi)}}\,\exp\Bigl[\,
  \int_0^{\beta\hbar} du\,{}^t\!\rho \,K(u)+\frac {1}{2}
  \int_0^{\beta\hbar} du \int_0^{\beta\hbar} dv\,
  {}^t\!K(u)\,\Phi(u-v)\, K(v)\,\Bigr]\,.
\label{zeta0}
\end{equation}
In Eq.(\ref{zeta0}) we have introduced the $2\times 2$ matrix $\Phi_{k\ell}(u-v)$
with elements
$\Phi_{11}(u-v)=m^2\omega^2\,\Phi_{22}(u-v)=m^2\omega^2\Lambda_f(u-v)\,$
and $\Phi_{12}(u-v)=-\Phi_{21}(u-v)=\Gamma_f(u-v)\,,$ where
\begin{eqnarray}
  & &\Lambda_f(u-v)=\frac{\hbar}{2m\omega\sinh
    f}\Bigl[\,\cosh(\,|\omega(u-v)|-f\,)-\frac{\sinh f}f\,\Bigr] \cr &
  &\cr & &\Gamma_f(u-v)=im\, \frac{\partial\Lambda_f(u-v)}{\partial v}
  = -\frac{i \hbar}{2}\frac{\sinh(\,|\omega(u-v)|-f\,)}{\sinh
    f}\,\Bigl[\, \theta(u-v) -\,\theta(v-u)\,\Bigr]\,.
\end{eqnarray}
While $\Lambda_f$ is always well defined, the value of $\Gamma_f$ for
$u=v$ is determined only if we specify the limit $v-u\rightarrow
0\,^\pm$ that must be taken, reflecting the commutation relation of
$\hat x$ and $\hat p$ at the same time. Moreover, a squared length
$~\alpha=\hbar/(2m\omega)\, ({\rm coth}f - 1/f)~$ and an energy $~E_a=
m \omega^2\alpha~$ are naturally defined by the parameters of the
system. It is also worth observing that
$\Lambda_f(0)=\Lambda_f(\beta\hbar)=\alpha\,$ and that $\Lambda_f(u-v)$ and
$\Gamma_f(u-v)$ have a vanishing average in $[0,\beta\hbar]$.

If we define the two-component vectors $\hat z={}^t( \hat p, \hat
x)$ and $y=\big(y_1(u),y_2(u)\big)$, after a lengthy but
straightforward calculation, the following general formula can be
derived in the low-coupling approximation \cite{CGTVV95}, {\it i.e.}
$\omega$ independent of $\eta$ and $\xi$:
\begin{eqnarray}
  & & \Bigl\langle{{\cal T}_u}\Bigl[\prod_{\nu=1}^N \hat F_{\nu}( \hat
  z_{i_\nu}(u_\nu))\Bigr]\Bigr\rangle\, =\,{\cal N}\prod_{\nu=1}^N
  F_{\nu}\Bigl(\frac \delta{\delta
    K_{i_\nu}}\Bigr)Z_0[K]\Big|_{{}_{K=0}}\, =\,{\cal N}\,\int
  \frac{d\eta\, d\xi}{2\pi\hbar}\, e^{\displaystyle{-\beta{\cal
        H}_{{}_{\rm eff}}(\eta,\xi)}}\, \Bigl\langle\Bigl\langle
  \prod_{\nu=1}^N
  F_{\nu}\big(\rho_{i_\nu}+y_{i_\nu}(u_\nu)\big)\Bigr\rangle\Bigr\rangle\,.
\label{genav}
\end{eqnarray}
Here, ${\cal N}=Z^{-1}_0[0]$ is the normalizing factor, and the
 double braces denote the average over the Gaussian
distribution of the variables $y_{i_\nu}(u_\nu)$, defined by the
second moments
\begin{equation}
\bigl\langle\bigl\langle y_i(v) y_j(u)
\bigr\rangle\bigr\rangle=\Phi_{ij}(u-v)\, .\label{gau}
\end{equation}
\noindent

The expression (\ref{genav}) with the definition (\ref{gau})
represents the main result of our paper. Complicated static Mori
products, {\it i.e.} moments of any order, can be evaluated by this last
equation. Static correlations are also obtained performing the appropriate
limit $u-v\rightarrow 0$.

In order to proceed on our path, we first look for an explicit
expression for averages of two functions separately depending on
momentum or displacement.  A direct calculation leads from Eq.
(\ref{genav}) to the following expansion:
\begin{eqnarray}
  & &\Bigl\langle{{\cal T}_2u}\Bigl[\hat A(z_i(v))\hat
  B(z_j(u))\Bigr]\Bigr\rangle\, =\,{\cal N}\,\int \frac{d\eta\,
    d\xi}{2\pi\hbar}\, e^{\displaystyle{-\beta{\cal H}_{{}_{\rm
          eff}}(\eta,\xi)}}\, \int \frac{dy\, dz}{2\pi}\,
  \frac{A(y)\,B(z)} {[\Phi_{ii}\Phi_{jj}-\Phi_{ij}^2(u-v)]^{1/2}}\cr &
  &\phantom{XXXXXXXX} \exp\Bigl\{-\,\frac {\Phi_{ii}\Phi_{jj}}
  {2[\Phi_{ii}\Phi_{jj}-\Phi_{ij}^2(u-v)]}
  \Bigl[\frac{\bigl(y-\rho_i\bigr)^2}{\Phi_{ii}}
  +\frac{\bigl(z-\rho_j\bigr)^2}{\Phi_{jj}}- \bigl({y-\rho_i}\bigr)\,
  \frac{\Phi_{ij}(u-v)}{\Phi_{ii}\Phi_{jj}}\,
  \bigl(z-{\rho_j}\bigr)\Bigr]\,\Bigr\}\,.
\end{eqnarray}
with $\Phi_{kk}\equiv\Phi_{kk}(0)$. We note that, when static
quantities are approached, i.e. the two imaginary times become equal,
the particular ordering of operators depending on both $\hat p$ and
$\hat x$ has to be specified. Any of the two possible orders, $p$-$x$
or $x$-$p$, is obtained from the equation above by simply taking the
appropriate limit $v-u\rightarrow 0^\pm$.

The final step involved in the calculation of higher order moments is
the integral expression for the average of the imaginary time ordered
product of two operators ${\hat A}$ and ${\hat B}$ at two different
times $u$ and $v$, both depending on momenta and displacements.
Without loss in generality we can consider $\hat p$-left ordered
operators, so that ${\hat A}=\hat A_1(p(u^+))\hat A_2(x(u))$ and the
similar for ${\hat B}$. By specializing Eq.(11) to this situation and
with evident meaning of the notation, we obtain
\begin{eqnarray}
  & &\Bigl\langle{{\cal T}_u}\Bigl[\hat A(p(u),x(u))\hat
  B(p(v),x(v))\Bigr]\Bigr\rangle\, =\,\Bigl\langle{{\cal
      T}_u}\Bigl[\hat A_1(p(u^+))\hat A_2(x(u))\,\hat B_1(p(v^+))\hat
  B_2(x(v))\Bigr]\Bigr\rangle\, = \cr & & =\,{\cal N}\,\int
  \frac{d\eta\, d\xi}{2\pi\hbar}\, e^{\displaystyle{-\beta{\cal
        H}_{{}_{\rm eff}}(\eta,\xi)}}\,
  \int\,d^4y\,\frac{A_1(y_1)\,A_2(y_2)\,B_1(y_3)
    \,B_2(y_4)}{(2\pi)^2\,{\det}^{1/2}(\Psi_{ij})}\,
  \exp\Bigl[\,-\frac12\sum\limits_{ij}(y_i-\rho_i)\,
  \Psi^{-1}_{ij}\,(y_j-\rho_j)\Bigr]
  \,,
\label{finale}
\end{eqnarray}
where now $\rho={}^t(\eta,\xi,\eta,\xi)$ is a four-component vector
and we have introduced the $4\times 4$ matrix $\Psi$ with elements
$\Psi_{11}=m^2\omega^2\Psi_{22}=\Psi_{33}=m^2\omega^2
\Psi_{44}=m^2\omega^2\alpha$,
$\Psi_{12}=\Psi_{21}=\Psi_{34}=\Psi_{43}=-i\hbar/2$,
$\Psi_{13}=m^2\omega^2\Psi_{24}=\Psi_{31}=m^2\omega^2
\Psi_{42}=m^2\omega^2\Lambda_f(u-v)$
and $\Psi_{14}=-\Psi_{23}=-\Psi_{32}=\Psi_{41}=\Gamma_f(u-v)$.

From Eq.(\ref{finale}), when $(u-v)\to 0^+$, we find the static average:
\begin{equation}
  \langle \hat A\hat B\rangle= {\cal N}\,\int \frac{d\eta\,
    d\xi}{2\pi\hbar}\, \langle\langle
  AB\rangle\rangle\,e^{\displaystyle{-\beta {\cal H}_{{}_{\rm
          eff}}(\eta,\xi)}}\,,
\label{statAB}
\end{equation}
indeed recovering our previous
result\cite{CTVVmagall,CGTVV95,Weylnote}.

As far as the Mori product is concerned, the well known series
expansion $(2\pi\epsilon)^{-1/2}
\,\exp\{-x^2/(2\epsilon)\}=\sum_{n=0}^\infty\,(1/n!)\,
(\epsilon/2)^n\,\delta^{(2n)}(x)\,$ appears to be an efficient tool to
approximate the static Mori product of general operators, as well as
their dynamical correlations, when the scales of the quantum
fluctuations in the system, ruled by $\hbar$ and the natural length
scale $\alpha$, are small.  Indeed, as the averages of $\Lambda_f(u)$
and $\Gamma_f(u)$ in $[0,\beta\hbar]$ vanish, we have
\begin{eqnarray}
  (\hat A(p,x)|\hat B(p,x)) && ={\cal N}\,(\beta\hbar)\, \int
  \frac{d\eta\, d\xi}{2\pi\hbar}\,e^{\displaystyle{-\beta{\cal
        H}_{{}_{\rm eff}}(\eta,\xi)}}\Big\{\,\, \langle\langle
  A(\eta,\xi)\rangle\rangle\,\langle\langle B(\eta,\xi)\rangle\rangle
  \,+ \cr && \!\!\hskip -0.6 truecm {1\over
    2}\Big[(m^2\omega^4A_{pp}B_{pp} +2
  m^2\omega^2A_{pq}B_{pq}+A_{qq}B_{qq})\sigma_f + (A_{pp}B_{qq}- 2
  A_{pq}B_{pq} +A_{qq}B_{pp})\mu_f\Big] \,+o(\alpha^2,\hbar^2)\Big\}\,
  ,
\label{morexp}
\end{eqnarray}
where the subscripts of $A$ and $B$ denote the derivatives, while
$\sigma_f$ and $\mu_f$ are the averages of $\Lambda_f^2$ and
$\Gamma_f^2$ in $[0,\beta\hbar]$. Therefore, as $\sigma_f$ is of order
$\alpha^2$ and $\mu_f$ of order $\hbar^2$, at the lowest order $(\hat
A(p,x)|\hat B(p,x))$ reduces to the ``classical like'' average of the
product of the Gaussian spreads of the two operators taken at the same
order:
\begin{equation}
  (\hat A(p,x)|\hat B(p,x)) ={\cal N}\,\beta\hbar\, \int \frac{d\eta\,
    d\xi}{2\pi\hbar}\, \langle\langle
  A(\eta,\xi)\rangle\rangle\,\langle\langle
  B(\eta,\xi)\rangle\rangle\, e^{\displaystyle{-\beta{\cal H}_{{}_{\rm
          eff}}(\eta,\xi)}}\,+o(\alpha,\hbar)\, .
\label{morexpsimpl}
\end{equation}

In the following discussion, for the sake of simplicity, we consider a
particle in a potential with a standard Hamiltonian
\begin{equation}
{\cal H}(p,x)=p^2/2m+V(x)\,;\:~~~ \:\:\:V(x)- V''(x_0)
(x-x_0)^2/2\equiv gv(x).
\end{equation}
where $x_0$ is the absolute minimum of the potential, and the quantum
coupling constant $g$ has been defined. The effective Hamiltonian is
therefore ${\cal H}_{{}_{\rm eff}}(\eta,\xi)=\eta^2/( 2m) + V_{{}_{\rm
    eff}}(\xi)$, where $m$ is now the particle mass, i.e. a constant,
and $V_{{}_{\rm eff}}(\xi)=\langle\langle
V(\xi)\rangle\rangle+\beta^{-1}\,\ln(f^{-1}\,\sinh f)$\,.

The simplest physically relevant quantity that can be discussed is given
by the imaginary time ordered correlation of two dynamical variables
$A(x)$ and $B(x)$, dependent upon the coordinates only.  For standard
Hamiltonians the integration over momentum can be done and we get the
following explicit expression for the correlation function:
\begin{eqnarray}
  {} & &\langle{{\cal T}_u}\Bigl[\hat A(x(v))\hat
  B(x(u))\Bigr]\rangle\, =\,{\cal N} \sqrt{\frac
    m{2\pi\beta\hbar^2}}\, \int d\xi\,e^{\displaystyle{-\beta
      V_{{}_{\rm eff}}(\xi)}} \,\int dy\,\int dz\cr {}& &
  \phantom{XXXXXXX}2\,A(y)\,B(z)\,\frac{\exp\Bigl\{\Bigr.-\displaystyle {(y-z)^2\over {4[\alpha-
      \Lambda_f(u-v)]}}\Bigl.\Bigr\}}
  {\phantom{XXX}\Bigl[4\pi(\alpha-\Lambda_f(u-v))\Bigr]^{1/2}}
  ~\frac{\exp\Bigl\{\Bigr.-\displaystyle{(y+z-2\xi)^2\over 4[\alpha+\Lambda_f(u-v)]}\Bigl.\Bigr\}}
  {\phantom{XXX}\Bigl[4\pi(\alpha+\Lambda_f(u-v))\Bigr]^{1/2}}\,.
\end{eqnarray}
\noindent
From the last equation, or by specializing equations (\ref{finale})
and (\ref{morexp}), the expansion for the Mori product is finally
obtained:
\begin{eqnarray}
  (\hat A(x)|\hat B(x)) &=&\,{\cal N}\,\beta\hbar\,\sqrt{\frac
    m{2\pi\beta\hbar^2}}\, \int d\xi\,e^{\displaystyle{-\beta
      V_{{}_{\rm eff}}(\xi)}} \Bigl\{\,A(\xi)B(\xi)+
  \frac{\alpha}2\,\Bigl[A''(\xi)B(\xi)+A(\xi)B''(\xi)\Bigr]\cr {}& &
  ~~~~~~~+\frac{\alpha^2}{8}\Bigl[A^{(iv)}(\xi)B(\xi)+
  +A(\xi)B^{(iv)}(\xi)+\Bigl(\,2+\frac {4 \sigma_f}{\alpha^2}
  \Bigr)\,A''(\xi)B''(\xi)\Bigr]\,\Bigr\}+o(\alpha^2)\,.\label{product}
\end{eqnarray}
Notice that for the displacement-displacement correlation function, as
probed by neutron scattering, only the first term in the last equation
(or in Eq. (\ref{morexp})) survives, and gives the total contribution
to $(\hat A(x)|\hat B(x))$ in the effective potential approach.
Comparing with Eq.(\ref{statAB}), we have $\langle \hat x^2\rangle =
(\beta\hbar)^{-1}(\hat x|\hat x) + \alpha\,$. This shows the twofold contribution of
the quantum fluctuations, {\it i.e.} the spreads of the potential and
of the quantity under averaging. The Mori product takes into account
the spread of the potential only. A result similar to our
Eq.(\ref{morexp}) was proposed in \cite{CaoVall} with {\it ad hoc}
assumptions and using only the first cumulant. As we can calculate the
zero moment, $\langle\omega\rangle^{(0)}=(x|x)$, and since the higher
moments are given by the other previous equations, the dynamic
correlations $\langle x(t)x\rangle$ can be eventually determined by
the continued fraction expansion
\cite{CGTVV95,Moriboth,CMGTVall,LoveseyM72}.

We like to conclude with some remarks on another possible dynamical
extension of our theory. We can assume that our system evolves with
the effective Hamiltonian as found for the thermodynamic behaviour.
This means that we take the commutator for its quadratic part, while
we consider the Poisson brackets for the one-loop-renormalized 
nonlinear part. In order to do this, we have to use a representation of
the operators describing the dynamical variables which permits to
unify quantum and classical dynamics, as, for instance, the
Weyl-representation \cite{CTVVmagall,CGTVV95} or the Liouvillian
scheme \cite{Moriboth}. To better understand such point we recall in
the first place that harmonic oscillators evolve by the same law both
in classical and in quantum dynamics: the differences between quantum
and classical statistical evolution are due to the thermal occupation
numbers, that are static quantities.  For real systems, the quantum
deviations from the harmonic behaviour are ruled by the coupling
constant $g$, related to quantum non linearity. If $g$ is vanishing,
the non commutativity of operators at different times can be neglected
and Eq.(\ref{morexp}) can be assumed to hold its validity at different
real times, provided also that $\alpha$ is small enough, so that the
Gaussian spreads of $\hat A$ and $\hat B$ do not overlap. At this
level, the averages in time of the quantities $\langle\langle {\hat
  A}\rangle\rangle$ and $\langle\langle {\hat B}(t)\rangle\rangle$,
evolving with the effective potential $V_{{}_{\rm eff}}(\xi)$
\cite{CaoVall,MartynaCao}, can provide an approximation for the
time-dependent Mori product $({\hat A}|{\hat B}(t))$. Finally its
Fourier transform gives then the relaxation function
$R_{A,B}(\omega)$, that is connected to the Fourier transform
$C_{A,B}(\omega)$ of the dynamical correlation function $\langle{\hat
  A}{\hat B}(t)\rangle$ by means of the fluctuation-dissipation
theorem , as in Eq.(2). This last operation simply restores the
correct static quantum occupation numbers.

We stress that the results become exact when $g\to 0$, {\it i.e.} for
quantum harmonic oscillators with a classical nonlinear interaction
term. Of course, the other exact limit is the classical system. For
finite values of $g$, the validity of this scheme involves also the
amplitudes of the Gaussian fluctuations ruled by the parameter
$\alpha$. Therefore, there is the same behaviour found for approaching
static correlators with the effective Hamiltonian
\cite{CGTVV95,GTV88}. In that case, the lowest limit of temperature
was found to be related to both parameters $g$ and $\alpha$. For the
dynamic correlators, this procedure yields a good
approximation for times up to the order of $\hbar\beta$, for which the
use of the effective potential makes sense in the calculation of the static
quantities at lowest order, reproducing for instance a correct second
moment for the displacement-displacement dynamic correlator with a
well-behaved classical long time decay.


\begin{thebibliography}{10}
\bibitem{Weylnote} We recall that, as clearly observed in
  Ref. \protect{\cite{CGTVV95}}, the final result for the static
  average is independent of the particular ordering (representation)
  chosen for a given operator. In fact, different ordering correspond to
  different moments of the gaussian spread, which lead to the same
  final result.
\end{thebibliography}
\end{document}